\documentclass[conference]{IEEEtran}
\IEEEoverridecommandlockouts
\usepackage{cite}
\usepackage{amsmath,amssymb,amsfonts}
\usepackage{algorithmic}
\usepackage{graphicx}
\usepackage{hyperref}
\usepackage{textcomp}
\usepackage{xcolor}
\usepackage{url}
\def\BibTeX{{\rm B\kern-.05em{\sc i\kern-.025em b}\kern-.08em
    T\kern-.1667em\lower.7ex\hbox{E}\kern-.125emX}}
\begin{document}

\title{Power Consumption of Video-Decoders on Various Android Devices\\
\thanks{This work was partially supported by Russian Foundation for Basic Research under Grant 19-01-00785 a.}
}

\author{\IEEEauthorblockN{Roman Kazantsev and Dmitriy Vatolin}
\IEEEauthorblockA{\textit{The Faculty of Computational Mathematics and Cybernetics} \\
\textit{Lomonosov Moscow State University}\\
Moscow, Russia \\
\{roman.kazantsev,dmitriy\}@graphics.cs.msu.ru}
}

\maketitle

\begin{abstract}
The critical constraint of mobile devices is a limited battery life that is significantly reduced during video playback. The power efficiency of video playback mainly depends on the used compression standard, video-decoder, and device model. We propose a software-based method to estimate the power consumption of video-decoders on various Android devices. Experiments on two devices of the same model show a small variation of the power playback consumption and a lack of dependence between the power consumption and the battery level. We have implemented an automatic system that includes the VEQE Android application to measure the power consumption of decoders and a server to collect the power metrics. Our system has collected power-consumption and decoding-speed dataset for video-decoders of six standards (AV1, HEVC, VP9, H.264, VP8, and MPEG-4) operating on 285 devices, representing 147 models. We demonstrate some slices of the created dataset: the top 30 models and video-decoders in terms of power efficiency for playback and for decoding only, as well as video-decoder ratings by power consumption and decoding speed for a given device model.
\end{abstract}


\section{Introduction}
The number of people who watch video on a mobile device has grown constantly in recent years.  In general,  video-service providers must compress their videos and maintain high visual quality by  selecting  an  optimal  video codec  along  with  an appropriate  preset, in addition to using a suitable bitrate for smooth transmission. But  mobile devices have a critical constraint: limited battery life. Video services targeting these devices must therefore employ compression to save power yet still provide a rich user experience.

Video-decoding is a major drain on the battery during playback, so its power efficiency must be maximized. This feat is achievable through a two-step process:  First,  select a power-efficient video-compression standard and video-decoder for the device. Second, find an encoding preset that ensures power-efficient decoding of the resulting bitstream. In this work we present a dataset that can help execute the first step that is, selecting a power-efficient video-decoder for a given device. In most cases, mobile-device users would rather watch video in power-saving mode to extend battery life than watch it in high quality. Moreover, the small size of mobile-device displays means users are less able to discern visual-quality artifacts and shortcomings.

In this paper we propose a software-based system that automatically creates power-efficiency dataset for video-decoders running on various Android devices. The system consists of a client Android application to make power measurements and a server to collect the data from clients. Using this system, we created a dataset with 285 samples (representing 147 device models). We depict some slices of this dataset in charts that exhibit video-decoder ratings by power efficiency for one model and for a multimodel comparison. The dataset can aid in developing video-decoding-optimization methods by informing selection of a power-efficient video standard and video-decoder for a given device.

In the following sections we provide an overview of methods for video-decoding-energy minimization in related works, we describe our software-based method for estimating video-decoding energy, and we present our approach to creating and analyzing the dataset.

\section{Related Work}
Herglotz et al. \cite{JointOptRate} introduced Decoding-Energy-Rate-Distortion Optimization (DERDO), which extends the traditional rate-distortion optimization (RDO) but aims to encode video such that the decoding process is more power efficient. To estimate decoding energy, \cite{JointOptRate} and \cite{DecodingEnergyModeling} propose a model that uses the features of a bitstream to predict the power necessary to decode it. The model receives a bitstream-feature vector that includes Intra and Inter blocks, nonzero residual coefficients, transform-skip flags, and in-loop filter usage. In \cite{JointOptRate}, the authors reasonably claim that decoding power is not necessarily proportional to the bit-stream size because better compression often employs complex video-coding tools that necessitate more energy for decoding. Therefore, energy savings are possible at the expense of bitrate or video quality. The DERDO approach can save 5--17\% of energy at the expense of a 6--24\% bitrate loss in case of local playback. Online streaming slightly drops the energy savings.

Fernandes et al. \cite{GreenMetadata} described a system that encapsulates Green Metadata into a bitstream for power-efficient video playback. The metadata includes the characteristics of video content that needs to be decoded and displayed. The first metadata is RGB statistics of video frames that are used for adjusting RGB values to preserve visual quality in case of backlight reduction. The display's power savings can be performed on account of backlight reduction and varying RGB values changes the power consumption negligibly. The second one is complexity metrics that estimate the decoding complexity within some period and help to lower the CPU frequency with guarantee to complete the decoding within frame-rate deadlines. This approach passed the MPEG standardization but we have not found any industry video codec that implements it.

Sidaty et al. \cite{sw_hevc_decoder} proposed approximate computing methods to replace the HEVC standard's original motion-compensation (MC) and in-loop filters. The MC filter's approximation level defines several taps for the MC interpolation filters. The in-loop (deblocking and sample-adaptive-offset) filters introduce the skip-control parameter, which defines the frequency at which the in-loop filters are skipped. These approximate computing methods allow decoding-energy reductions of up to 20\% depending on the approximation level and the skip-control parameter. According to the authors, no significant quality degradation is noticeable on mobile devices, particularly in the intermediate-approximation configuration.

Yadav et al. \cite{PowerConsumptionAndroidDevices} proposed a software-based approach that uses a monitoring tool for power measurement. This tool reports the power usage of different Android device functions: the CPU, display, and Wi-Fi. These values enable computation of decoding energy through time as the video application launches. The tool provides accurate measurements, but because it only works on few Android devices, it lacks generality.

Sostaric et al. \cite{PowerConsumptionVideoDecoding} tested several video-decoders on Android devices using a hardware-based tool. They concluded that the best choice for mobile devices is to use older standard video codecs with advanced options. For example, MPEG-4 can outperform H.264 in a tradeoff among visual quality, power consumption and bitrate. Although the authors analyzed custom software codecs, modern devices employ a variety of hardware codecs, which are the primary choice for video applications.

Hu et al. \cite{EnergyAwareVideoStreaming} investigated methods to control a wireless interface and set the CPU frequency for decoding the next frame in a sequence. Such methods can reduce decoding energy by 9--17\%, whereas the frame-rate drop is less than 3\%.

The decoding-energy-optimization methods above require major modifications to video codecs and standards, and they are not hardware agnostic with regard to power measurements. In this paper we propose a way to create power-efficiency data set for hardware- and software-based video-decoders running on different Android devices. This dataset can aid in developing methods to predict which decoder is most power efficient for a certain device and resolution.

\section{Method for Estimating Video-Decoding Energy}
We propose a software-based approach to measuring the power consumption of video-decoders on Android devices. Our approach is generic, meaning it is suitable for all device models running Android 5.0 or higher. Android 5.0 was introduced in 2014. Its main idea is to retrieve the device’s battery level using the Android API’s BatteryManager feature. When decoding video, our method constantly queries the battery level (the current battery-charge percentage) using the Android API, logging that level and a time step in case of a change. The Android API also makes it easy to get other device parameters: the voltage; battery capacity, type and health; device model and manufacturer; and more.

The main problem with our proposed approach is that the battery level rarely changes, so it must decode the same bitstream multiple times until the battery level changes. For greater accuracy, we wait for this level to drop by 3\%. The device runs in autonomous mode during these measurements. To ensure more-accurate measurements, we track the iterations and frame indices at which the battery level changes.  
Assume the levels are $B_0$ and $B_1$ at times $T_0$ and $T_1$ (in seconds), 
respectively, corresponding to iterations $N_0$ and $N_1$ and frame indices $n_0$ and $n_1$. Also assume the sequence contains $n_{seq}$ frames. The formula for computing the relative battery consumption (\%) when decoding the entire bitstream and the formula for computing the average decoding speed (frames per second) are as follows:
\begin{equation}
\Delta_{seq} = \frac{B_0 - B_1}{(N_1 - N_0)n_{seq} + n_1 - n_0} \cdot n_{seq},
\end{equation}
\begin{equation}
\upsilon = \frac{(N_1 - N_0)n_{seq} + n_1 - n_0}{T_1 - T_0}.
\end{equation}

In this work we compute two valuable power metrics. The first is relative battery consumption per hour of playback $\Delta_{play}$ (in \% per hour), which lets end-users estimate how long they can watch video in autonomous mode. The second is video-decoding energy per hour, $\Delta_{decode}$ (in mA), which helps Android-device manufacturers estimate the power efficiency of hardware video-decoders. Note that $\Delta_{seq}$ includes the energy that the display consumes. We define the display's power consumption per hour as $\Delta_{screen}$ (\% per hour) and compute it using the same approach by which we track battery-level changes on a device in idle mode with its display on. We calculate $\Delta_{play}$ as follows:
\begin{equation}
\Delta_{play} = \Delta_{seq} \cdot \frac{fps}{n} \cdot 3600 + \Delta_{screen} \max \left(0; 1 - \frac{fps}{\upsilon}\right),
\label{play}
\end{equation}
where $n$ is the number of frames in the sequence and $fps$ is the number of frames per second required for playback.
Equation \eqref{play} includes a term $ \Delta_{screen} \max \left(0; 1 - \frac{fps}{\upsilon}\right)$ compensating the energy that the display consumes for some time
when decoding is complete in case of $\upsilon > fps$. Note that $\upsilon \geq fps$ is required for normal playback.
The formula to compute $\Delta_{decode}$ (mA) is the following:
\begin{equation}
\Delta_{decode} = V \left(\Delta_{seq} \cdot \frac{\upsilon}{n} \cdot 3600 - \Delta_{screen} \right),
\end{equation}
where $V$ is a battery capacity (mA $\cdot$ h). Both metrics, $\Delta_{play}$ and $\Delta_{decode}$, can help to benchmark mobile devices and video-decoders.

Android devices comprise multiple video-decoders implementing different compression standards (AV1, HEVC, VP9, H.264,  VP8, and MPEG-4). These decoders divide into hardware, software and hybrid types. To estimate the power consumption, we selected three video sequences of SD, HD and Full HD resolution with the frame-rate set to 25 fps, then encoded them for each compression standard using FFmpeg with the medium bitrates. The video sequences were encoded for the streaming case with a fixed-cadence group of picture (GOP), GOP-length equal to two seconds, three references for motion compensation, two B-frames, and maximum bitrate equal to 1.2 of the average bitrate.
The settings restrict to use the Main Profile for H.264 and HEVC with Level 4.2 that supports selected resolutions, frame-rate, and maximum bitrate on all Android 5.0 and higher devices. Our choice of video sequences is based on their distinctive spatial and temporal complexities. Defining the averages of these complexities are the mean spatial perceptual information, $SI_{mean}$, and the mean temporal perceptual information, $TI_{mean}$ \cite{SubVideoQual}. Table~\ref{tabvideos} lists the $SI_{mean}$, $TI_{mean}$ and the average bitrate for each video.

\begin{figure}[htbp]
\centerline{\includegraphics[width=9.0cm]{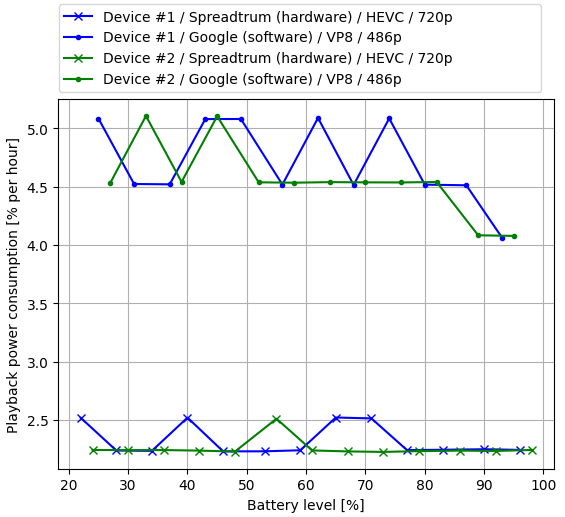}}
\caption{Playback power consumption, $\Delta_{play}$, of two ZTE Blade A5 2019 devices and its dependence on the battery level.}
\label{multidevice}
\end{figure}

Before employing the method in practice we arrange an experiment to check how the playback power consumption varies on different devices of the same model and the dependence of the power consumption on the battery level. The results for two ZTE Blade A5 2019 devices are shown in Fig.~\ref{multidevice}. We receive the standard deviation for each device equal to 0.5\% and the absolute difference between the power metric means on the devices equal to 0.01\%.

Adequate power estimation using our method must satisfy the following requirements:
\begin{enumerate}
\item The device must operate in autonomous mode;
\item The device must have neither too low a battery level (less than 20\%), which allows the device to run in safe-power mode, nor too high a level (greater than 95\%), which makes the battery more resistant to discharging;
\item Power estimation for decoding one bitstream must expend at least 3\% of the battery capacity;
\item Power measurements may take place in different charge cycles, because devices can have more than 20 video-decoders, and testing all of them at three resolutions requires at least 180\% of the battery's capacity.
\end{enumerate}

We implemented this method in the VEQE Android application, which is available from Google Play: \\
\url{https://play.google.com/store/apps/details?id=ru.msu.cs.graphics.veqe_mobile}.

\section{Dataset Creation}
We developed an automatic system to collect power-efficiency data for video-decoders in different Android devices. It consists of the VEQE application, a client for testing, and a server for collecting data from all clients once the testing is complete. Since devices may implement multiple video-decoders, and since a full battery charge may be insufficient to test all of them, the application can report to the server intermediate results in case of incomplete testing or premature cancellation, and it can save a checkpoint from which it can continue testing after the client recharges. Aside from power metrics, the application also estimates decoding speed for every video-decoder.

To create the dataset we used Yandex.Toloka, a cloud-based crowdsourcing platform: we collected and labeled extensive data by posting an exercise that paid participants to launch our application. To retain good-quality data submissions and discard fake ones, we prepared a server-side script that helps avoid duplicates in the dataset by checking device properties (a serial number and a build host). The script also ensures the samples are complete --- that is, all available video-decoders on a given device undergo testing --- by reporting the ratio of video-decoders tested for each submission. It allows validation of samples in addition to uploading the dataset semiautomatically --- a necessary feature because of the myriad submissions.

Finally, we received data from 285 devices representing 147 models and 60 video-decoders (43 hardware, 17 software and 1 hybrid). The next section describes processing and analyzing the dataset.

\section{Data Processing and Evaluation}
The data we collected is raw and needs further processing to exclude anomalies and to compute $\Delta_{decode}$ as well as $\Delta_{play}$ for each video-decoder and resolution. 
If certain samples correspond to the same device model, we combined them and computed the mean $\Delta_{decode}$, $\Delta_{play}$ and $\upsilon$ values for each decoder
and resolution of a given model.

\begin{table}[htbp]
\caption{Description of videosequences}
\begin{center}
{
\begin{tabular}{|l|l|l|l|l|}
\hline
\textbf{Video}        & \textbf{Resolution}           & \textbf{$SI_{mean}$} & \textbf{$TI_{mean}$} & \textbf{Bitrate (Kbps)} \\ \hline
\textbf{Shakewalk} & $640\times480$               & 0.058                          & 124.76                        & 2560                             \\ \hline
\textbf{Tractor}     & $1280\times720$              & 0.071                          & 100.57                        & 5120                              \\ \hline
\textbf{Zombie}     & $1920\times1080$            & 0.073                          & 104.66                        & 12288                            \\ \hline
\end{tabular}
}
\label{tabvideos}
\end{center}
\end{table}

Some participants in the exercise we posted on the crowdsourcing platform generated low-quality data if they used a device that was powered through an AC adapter or had a locked screen during the benchmarking. The number of devices of the same model is small and varies up to nine in our dataset so statistical methods turned out rough for anomaly detection. So we manually walked through each model with a suspicious deviation for the power metrics and removed obvious outliers. We detected 8\% anomalies and low-quality instances.

After cleaning the dataset and computing $\Delta_{play}$ as well as $\Delta_{decode}$ we have done analysis of the data by collecting different statistics and we conclude the following:
\begin{itemize}
  \item Hardware decoders surpass software decoders considerably in power efficiency and decoding speed, particularly for high resolutions. Software decoders win only in 2--6\% cases depending on resolution.
  \item Video-decoders of MPEG-4 (in 29--32\% cases) and H.264 (in 25--31\% cases) standards outperform decoders of other standards in terms of power efficiency for all resolutions.
  \item Software decoders of MPEG-4, VP8 and AV1 standards can nevertheless remain efficient for low-resolution videos in 6\% cases.
  \item AV1 decoder presents only on devices with Android 11 where software AV1 decoder from Google is the most power-efficient for low resolutions in 38\% cases but it can be inapplicable to real-time decoding on some devices due to a low decoding speed.
\end{itemize}

\begin{figure}[htbp]
\centerline{\includegraphics[width=7.6cm]{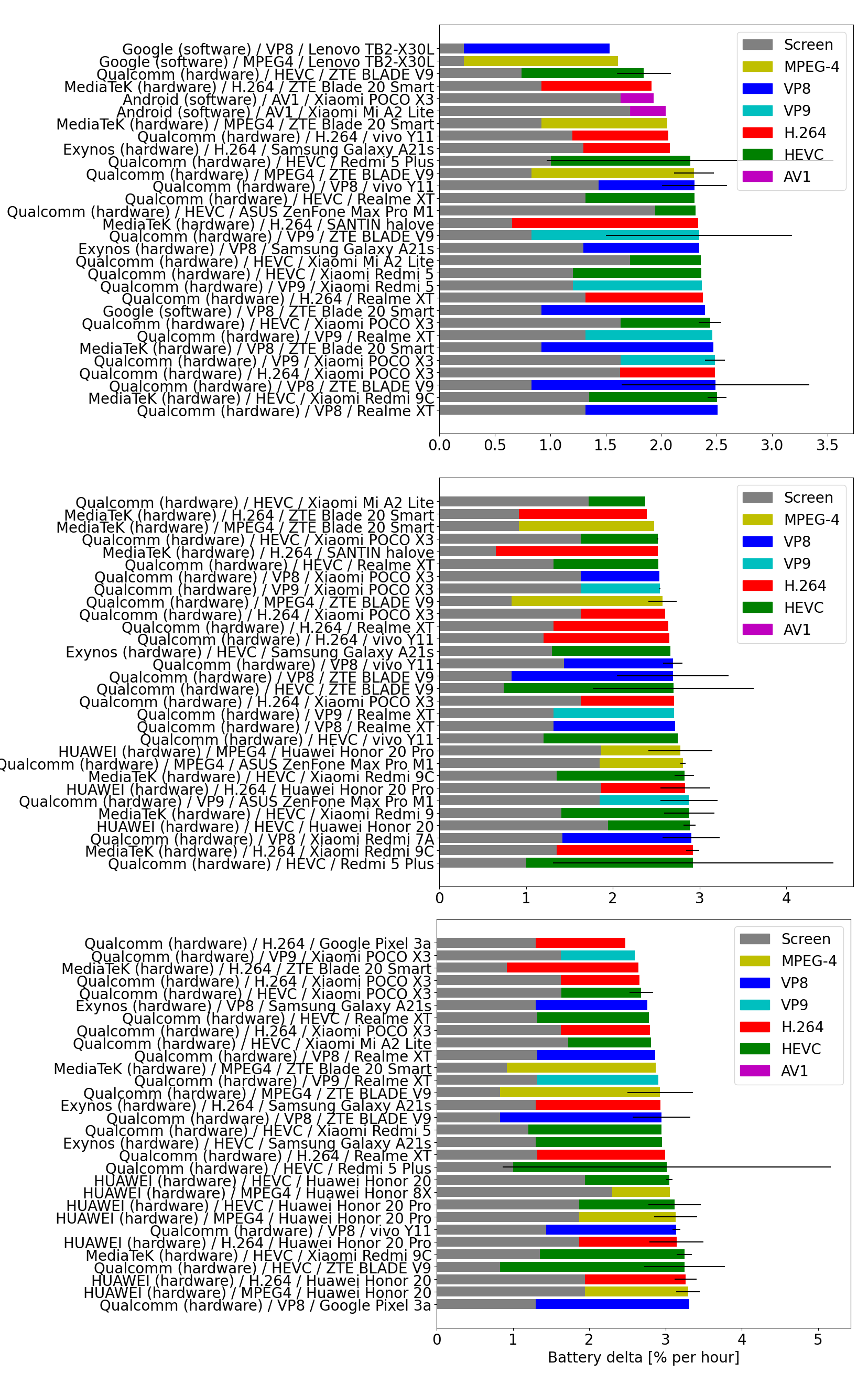}}
\caption{Playback power consumption, $\Delta_{play}$, for the top 30 video-decoders at SD, HD and Full HD resolution (top to bottom).
Horizontal errorbars correspond to standard deviations.}
\label{multimodel_playback}
\end{figure}

\begin{figure}[htbp]
\centerline{\includegraphics[width=7.6cm]{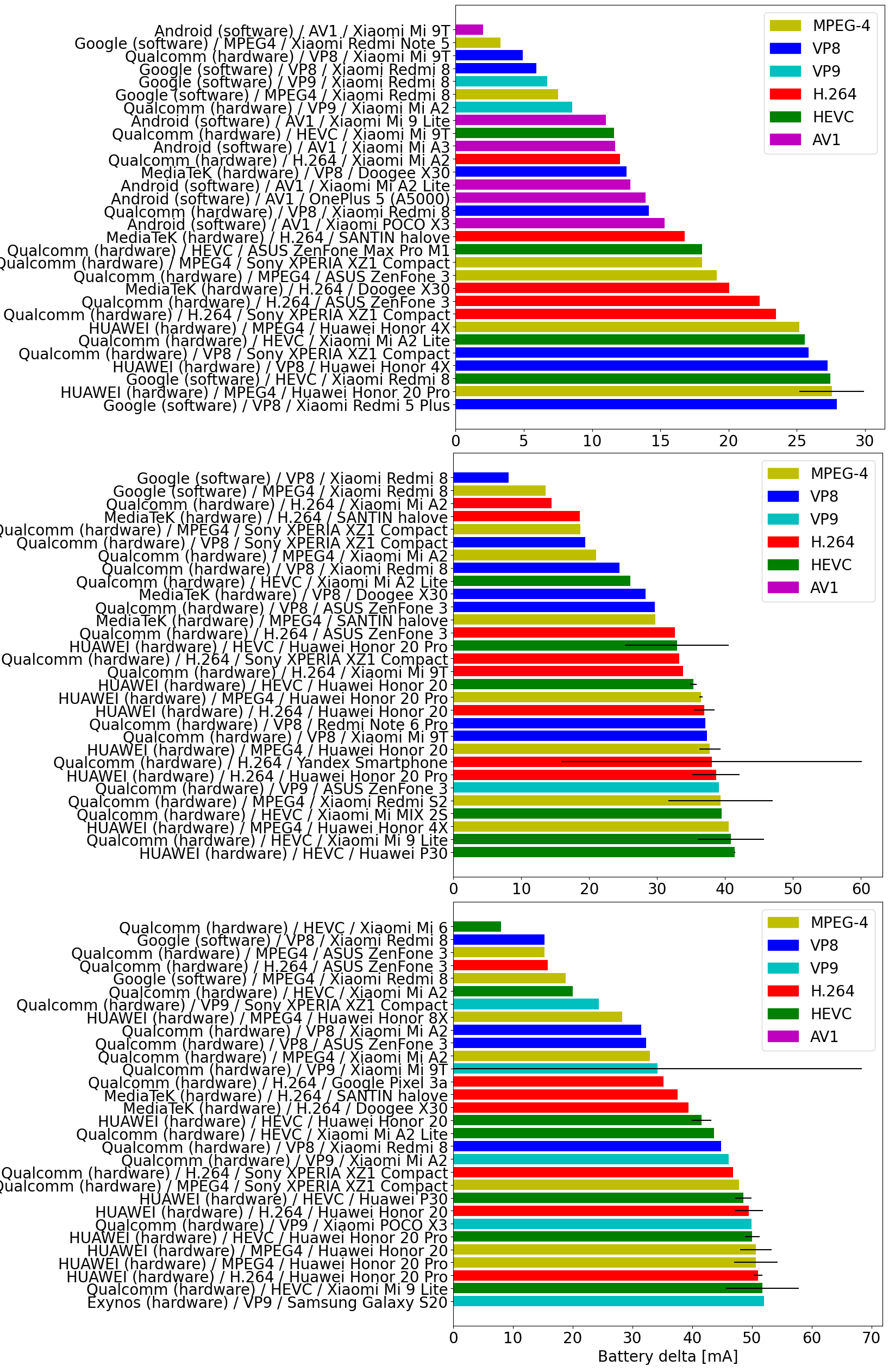}}
\caption{Power consumption, $\Delta_{decode}$, for the top 30 video-decoders at SD, HD and Full HD resolution (top to bottom).
Horizontal errorbars correspond to standard deviations.}
\label{multimodel_pure_decoding}
\end{figure}

We depict some slices of the dataset in charts to rate video-decoder power efficiency for a multimodel comparison in Fig.~\ref{multimodel_playback} and \ref{multimodel_pure_decoding}. The playback power consumption $\Delta_{play}$ can depend on battery capacity. For example, the large battery capacity helps Lenovo TB2-X30L Tablet accidentally hitting the dataset to bypass smartphones with half the capacity for SD resolution. However, for higher resolutions, this tablet model has inefficient video-decoders.
Most of the positions in both top 30 ratings are occupied with smartphones from Xiaomi and Huawei. Samsung and Google smartphones hardly fit these ratings despite their models are well-represented in the dataset.

\begin{figure}[htbp]
\centerline{\includegraphics[width=8.0cm]{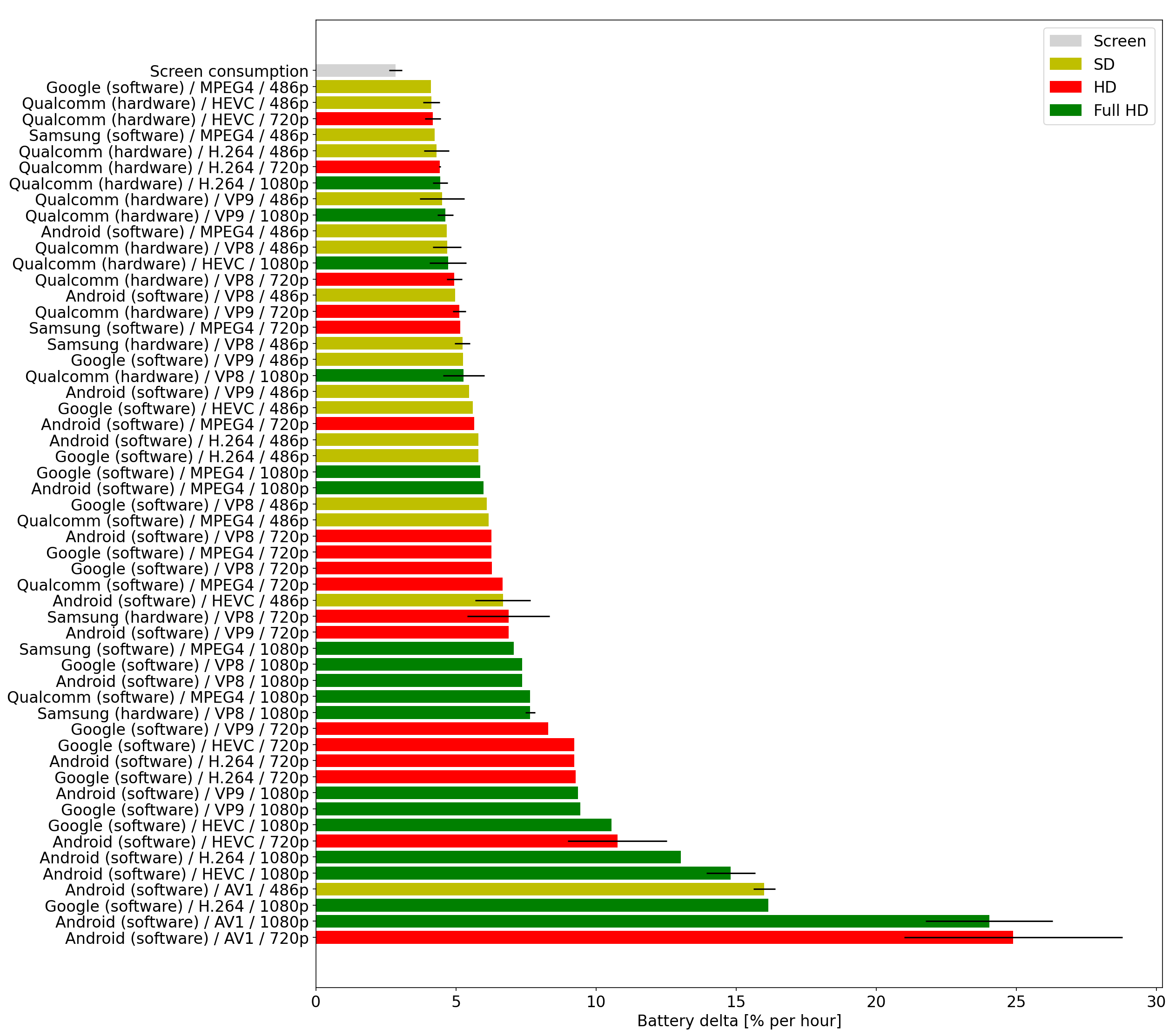}}
\caption{Playback power consumption of video-decoders, $\Delta_{play}$, running on a Samsung Galaxy A70 smartphone (all resolutions).
Horizontal errorbars correspond to standard deviations.}
\label{onemodel_playback}
\end{figure}

\begin{figure}[htbp]
\centerline{\includegraphics[width=8.0cm]{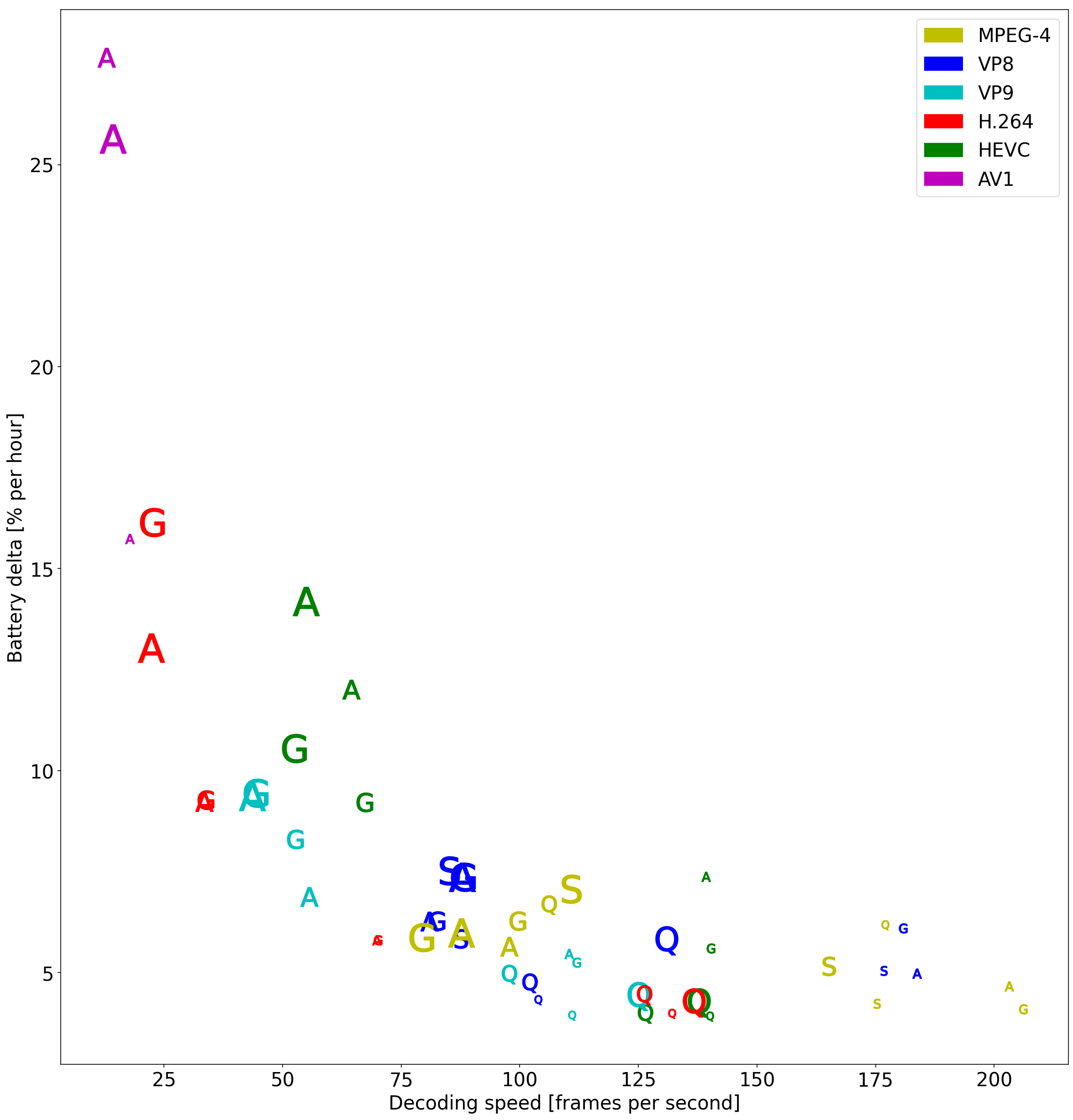}}
\caption{Playback power consumption, $\Delta_{play}$, versus decoding speed, $\upsilon$, on a Samsung Galaxy A70 (all resolutions). A=Android; G= Google; Q=Qualcomm; S=Samsung. Small letters indicate SD, medium letters indicate HD, and big letters indicate Full HD.}
\label{onemodel_power_vs_speed}
\end{figure}

Besides, we rate video-decoder power efficiency for a concrete model, Samsung Galaxy A70, in Fig.~\ref{onemodel_playback} and \ref{onemodel_power_vs_speed} where Qualcomm hardware decoders noticeably outstrip software decoders and AV1 decoder demonstrates unacceptable decoding speed.

\section{Conclusion}
In this work we proposed a software-based method to estimate the power consumption of video-decoders running on various Android devices. This method serves in the VEQE Android  application and allowed us to create a dataset using the power metrics from 285 Android devices, representing 147 models and 60 video-decoders. We tested video-decoders for six standards (MPEG-4, VP8, H.264, VP9, HEVC and AV1) and nine video-decoder manufacturers (Exynos, Google, Huawei, Imagination Technologies, Intel, MediaTek, Qualcomm, Samsung and Spreadtrum). Hardware decoders outperform software decoders in power efficiency and decoding speed, but software decoders of MPEG-4, VP8 and AV1 standards can remain efficient for low-resolutions. Due to a small number of devices that correspond to the same model in the dataset we did not manage to apply statistical methods for anomaly detection and did cleaning manually.

\section{Future Work}
As shown in Fig.~\ref{multimodel_playback} and \ref{multimodel_pure_decoding} we encounter a solid standard deviation in the power metrics for some models. To avoid low-quality data in the future, we plan to implement a protection mechanism against the use of smartphones in parallel to the VEQE application run. Also, we must exclude smartphones with old batteries that can affect the statistics and such phones can be detected using battery cycle count from $/sys/class/power\_supply/battery/cycle\_count$ file.

\end{document}